\documentclass[aip,rsi,reprint,graphicx]{revtex4-1}

\usepackage[utf8]{inputenc}
\usepackage[T1]{fontenc}
\usepackage{ae,aecompl}
\usepackage[english]{babel}
\usepackage{graphicx}
\usepackage{hyperref}
\usepackage{epstopdf}
\usepackage{color}
\usepackage{multirow}
\usepackage{natbib}

\begin{document}

\author{F. Arnold}
\affiliation{Max Planck Institute for Chemical Physics of Solids, 01187 Dresden, Germany}
\author{M. Naumann}
\affiliation{Max Planck Institute for Chemical Physics of Solids, 01187 Dresden, Germany}
\affiliation{Physik-Department, Technische Universit\"at M\"unchen, 85748 Garching, Germany}
\author{Th. L\"uhmann}
\affiliation{Max Planck Institute for Chemical Physics of Solids, 01187 Dresden, Germany}
\author{A. P. Mackenzie}
\affiliation{Max Planck Institute for Chemical Physics of Solids, 01187 Dresden, Germany}
\affiliation{School of Physics and Astronomy, University of St Andrews, St Andrews KY16 9SS, U.K}
\author{E. Hassinger}
\affiliation{Max Planck Institute for Chemical Physics of Solids, 01187 Dresden, Germany}
\affiliation{Physik-Department, Technische Universit\"at M\"unchen, 85748 Garching, Germany}

\title{Application of SQUIDs to low temperature and high magnetic field measurements - Ultra low noise torque magnetometry}

\date{\today}


\begin{abstract}
Torque magnetometry is a key method to measure the magnetic anisotropy and quantum oscillations in metals. In order to resolve quantum oscillations in sub-millimeter sized samples, piezo-electric micro-cantilevers were introduced. In the case of strongly correlated metals with large Fermi surfaces and high cyclotron masses, magnetic torque resolving powers in excess of $10^4$ are required at temperatures well below $1\,\mathrm{K}$ and magnetic fields beyond $10\,\mathrm{T}$. Here, we present a new broadband read-out scheme for piezo-electric micro-cantilevers via Wheatstone-type resistance measurements in magnetic fields up to $15\,\mathrm{T}$ and temperatures down to $200\,\mathrm{mK}$. By using a two-stage SQUID as null detector of a cold Wheatstone bridge, we were able to achieve a  magnetic moment resolution of $\Delta m= 4\times10^{-15}\,\mathrm{J/T}$ at maximal field and $700\,\mathrm{mK}$, outperforming conventional magnetometers by at least one order of magnitude in this temperature and magnetic field range. Exemplary de Haas-van Alphen measurement of a newly grown delafossite, PdRhO$_2$, were used to show the superior performance of our setup.

\end{abstract}

\maketitle

\section{Introduction}

\noindent A central issue of nowadays solid state physics is the down-scaling of sample size due to ever growing demands on purity and single crystallinity. This is particularly true for quantum oscillation studies of new strongly correlated electron systems, unconventional metals and superconductors \cite{Shoenberg,Lee06,Si10,Singleton10,Si16,Mackenzie17}. In some cases, single crystals of these materials can only be grown in a few micron to sub-millimeter size and are therefore impracticable for conventional solid state methods. In addition, magnetization and resistance resolving powers in excess of $10^4$ are required to observe de Haas-van Alphen or Shubnikov-de Haas oscillations in large Fermi surface metals.\\
In the past, highly sensitive ac-susceptometers \cite{Pobell,Amann17} and bronze-foil lever magnetometers \cite{Brooks87,Kampert} were used to measure magnetizations and magnetic anisotropies of bulk samples at millikelvin temperatures and in high magnetic fields, achieving resolutions of $\Delta m \approx 5\times10^{-13}\,\mathrm{J/T}$ \cite{Albert,Grundler}.\\
In the quest for ever smaller quantum oscillation signals more sensitive techniques and read-out schemes are required. Ultra low noise read-out schemes, using Complementary Metal-Oxide-Semiconductor (CMOS) and High-Electron-Mobility transistors (HEMT), were implemented for ultra high source impedance \cite{Libioulle03,Proctor15} and medium to very high frequency applications \cite{Oukhanski03,Robinson04} respectively. Although offering extremely high gain and bandwidth, their performance is limited by charge carrier freeze out at low temperatures, giving rise to shot and random telegraph noise. Thus low frequency MOS-based electronics are often stabilized at elevated temperatures around $100\,\mathrm{K}$, where temperature drift and the associated long term gain stability become an issue. Alternatively, low temperature transformers (LTTs) provide good temperature stability at liquid helium temperatures \cite{Pobell,CMR}. Their gain and bandwidth, however, strongly depend on the matched impedance either side of the transformer. Thus LTTs have mostly been applied to circuits with low source impedance.\\
In the preceding decades, Superconducting-Quantum Interference Devices (SQUIDs) became a new path to achieving highest signal-to noise ratios (SNRs) by virtual noiseless amplification of current signals \cite{Clarke,Drung07}, outperforming the hitherto known amplifiers and LTTs.\\
As a result, SQUIDs have been introduced to many low temperature applications such as resistance measurements \cite{Barnard78,Rowlands76,Romero89}, SQUID NMR \cite{Lusher98,Arnold14Pro}, MRI \cite{Espy13}, ESR \cite{Sakurai11}, microcalorimetry \cite{Mears97,Enss05,Kempf17} and Johnson noise thermometry \cite{Casey14,Rothfuss13} obtaining unprecedented precision. However, thus far, most of these techniques were restricted to zero or low magnetic fields, as SQUIDs are notoriously difficult to use in high magnetic fields.\\
More recently, with the introduction of superconducting shielding, high field resistance bridges as well as SQUID magnetometers were developed, extending the range of highly sensitive resistance \cite{Walker92,Barraclough} and magnetization measurements \cite{Bravin92,Nagendran11,QD} up to $14$ and $7\,\mathrm{T}$ respectively. Whilst SQUID magnetometers became a useful tool for quantum oscillation studies of macroscopic samples below $7\,\mathrm{T}$, SQUID resistance bridges suffered from excessive noise or could only be operated in static magnetic fields, making them impracticable for Shubnikov-de Haas experiments. Due to these technical limitations, neither of these techniques is suited for the study of microscopic samples of strongly correlated metals. To enable high precision magnetic measurements at high fields, piezo-electric micro-cantilever based magnetometers were introduced, measuring a sample's magnetic torque $\tau$ as the change of the piezo resistance \cite{Rossel96,McCollam11}. Here the magnetic torque $\tau=B\times m$, where $m$ is the sample's magnetic moment.\\
In this article, we report on the development of a new highly sensitive, high field, millikelvin SQUID torque magnetometer to measure 
quantum oscillations of sub-millimeter size samples. A two-stage dc-SQUID, located in the field compensated region of our cryostat, is utilized as an ultra-low noise current amplifier in a piezo-electric micro-cantilever Wheatstone bridge. Our setup achieves a hitherto unrivaled magnetization resolution of $\Delta m= 5\times10^{-15}\,\mathrm{J/T}$ at $15\,\mathrm{T}$ and $700\,\mathrm{mK}$ for the given temperature,magnetic field range and sample size\cite{Drung07,Rossel96}. The performance of our setup will be demonstrated by de Haas-van Alphen measurements of a newly grown PdRhO$_2$ delafossite single crystal \cite{Arnold17PRB}. Delafossites are correlated quasi-two dimensional electron systems with alternating triangular lattice layers of noble metal and transition metal oxide, showing exceptionally large electrical conductivities \cite{Hicks12,Hicks15,Kushwaha15,Mackenzie17}. We further compare our setup to conventional unamplified and LTT-amplified circuits, showing its superior resolving power.

\section{Experimental}

\subsection{General Description}

\begin{figure}[tb]
	\centering
		\includegraphics[width=0.95\columnwidth]{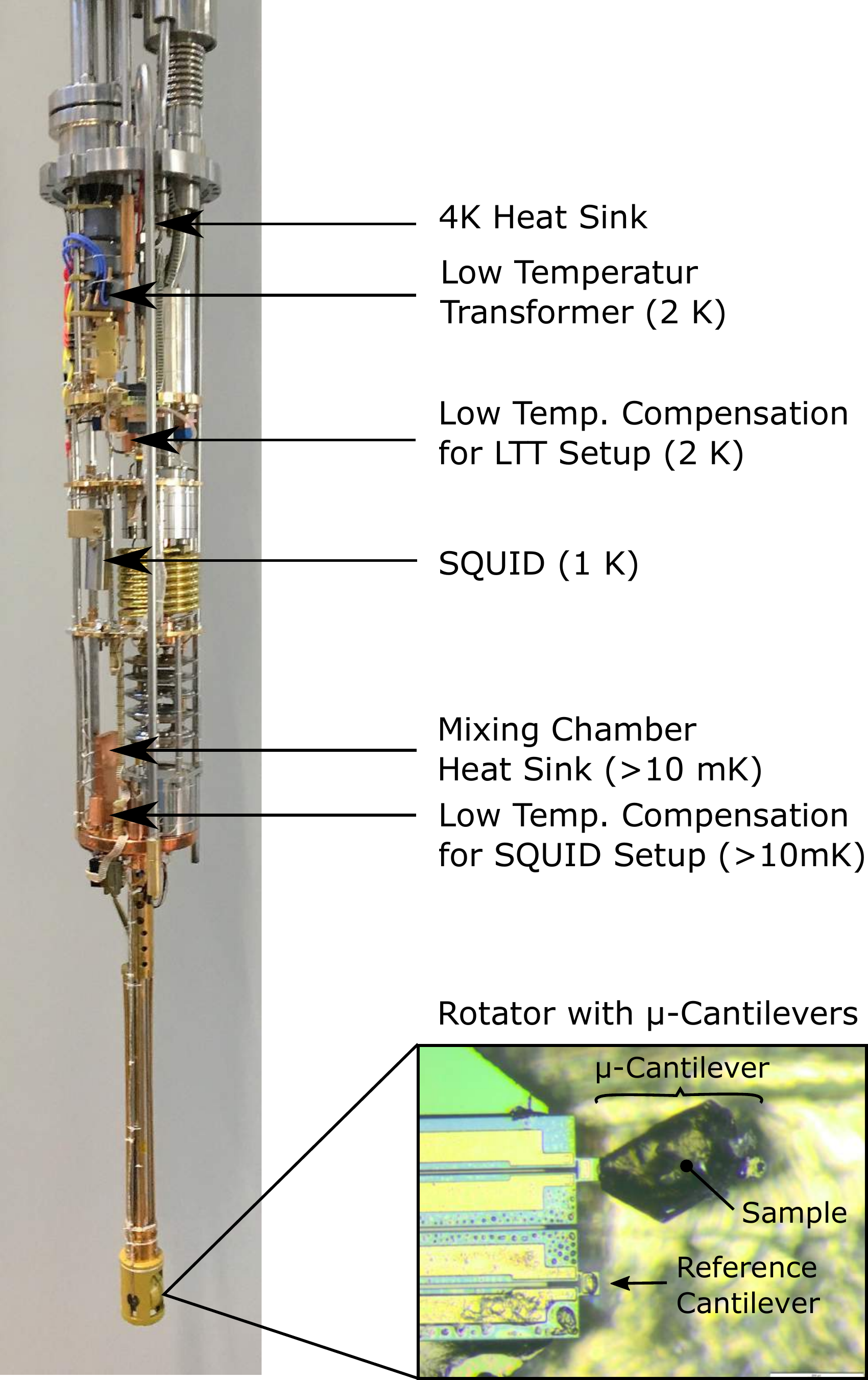}
	\caption{The graph shows an overview of the Oxford Instruments MX400 dilution cryostat insert. Arrows indicate the position of the individual components of the magnetic torque setup. The inset shows a zoom of the PRC400 piezo-resistive micro-cantilevers with the mounted PdRhO$_2$ sample.}
	\label{fig:Cryostat}
\end{figure}

\noindent An ultra-low noise, low temperature and high magnetic field torque magnetometer for sub-millimeter samples was built based on an Oxford Instruments MX400 dilution refrigerator, with a $15+2\,\mathrm{T}$ Nb$_3$Sn superconducting $\lambda$-stage magnet and a $270^\circ$ Swedish rotator (see Fig. \ref{fig:Cryostat}). The magnet of the cryostat is designed such to provide a field compensated region at the mixing chamber and low field region ($B<5\,\mathrm{mT}$ at full field) above. Self sensitive PRC400 piezo-resistive micro-cantilevers \cite{Hitachi} were used as magnetic torque sensors, which were mounted on a silver holder on the rotator. To improve sample thermalization, the back side of the micro-cantilevers was coated with gold and the rotator was thermally connected to the mixing chamber by an oxygen annealed silver wire braid. A calibrated RuO$_2$ thermometer was installed on the rotator for thermometry.\\
In the interest of achieving lowest noise levels all wiring is shielded in metal or superconductor capillaries. Superconducting wires and capillaries, that are $100\,\mu\mathrm{m}$ multi-filament CuNi-clad NbTi twisted pairs and tinned CuNi capillaries, were generally used in the field compensated and low field region of the cryostat. In the high field region between the mixing chamber and rotator, $75\,\mu\mathrm{m}$ copper twisted pairs shielded in an oxygen annealed industry-grade copper capillary were used instead. Here the annealed copper capillary acts as an almost perfect diamagnetic shield against low and high frequency fields, reducing pick-up noise from mechanical vibrations in high magnetic fields. For simplicity the micro-cantilevers and wiring on the rotator were unshielded. Capillaries and wires were heat sunk at the $1\,\mathrm{K}$-pot, still, cold plate and mixing chamber to reduce thermal leaks across the dilution unit. \\
Our setup uses a National Instruments $51.2\,\mathrm{kS/s}$, 24 bit PXIe-4463 signal generator and $204.8\,\mathrm{kS/s}$, 24 bit PXIe-4492 oscilloscope as the data acquisition system. Typical input noise levels of the PXIe-4492 oscilloscope are on the order of 5 to $10\,\mathrm{nV/}\sqrt\mathrm{Hz}$ in the frequency range between $10\,\mathrm{Hz}$ and $10\,\mathrm{kHz}$. A digital lock-in program was written to emulate a standard standalone lock-in amplifier on the PXI system. Its functionality includes measuring the in- and out-of phase components, phase angle, resistance as well as higher harmonics and power spectra of up to eight channels simultaneously.

\subsection{Conventional Room Temperature Balancing}

\noindent The magnetic torque exerted on the micro-cantilever is measured as a resistance change of the piezo-electric track implanted into the cantilever. The resistance of the piezo-electric track is usually measured in a Wheatstone bridge consisting of the sample and reference cantilever (see zoom display of Fig. \ref{fig:Cryostat}) as well as a room temperature potentiometer. The empty reference cantilever is used to compensate for the intrinsic temperature and magnetic field dependence of the track resistance.\\
Generally, room temperature compensated setups suffer from low frequency rf-noise, picked up outside the cryostat, comparably large input noise of room temperature amplifiers and AD-converters as well as Johnson noise of the balancing resistors. In our first measurements with the PRC400 cantilevers, this combination of factors set a noise floor, which limited the resolution of our measurement to $\Delta R/R \approx1.5\times10^{-5}$ (see discussion on the performance of the setup in Sec. \ref{sec:Performance}).

\subsection{Low Temperature Balancing}

\noindent In order to circumvent external noise sources and to boost the signal-to-noise ratio (SNR), low temperature amplification is desirable. For this a low temperature Wheatstone bridge must be implemented, balancing the micro-cantilever potential divider. In our case this cold compensation consisted of two high precision metal film SMD resistors, which were mounted in a shielded copper box to the $1\,\mathrm{K}$-pot (unamplified and LTT amplified setup) or mixing chamber (SQUID setup). In the low temperature transformer setup (section \ref{sec:LTT}) $R_\mathrm{C}=1\,\mathrm{k}\Omega$ resistors were used. Their shielding box was weakly thermally coupled to the $1\,\mathrm{K}$-pot to stabilize its temperature around $2\,\mathrm{K}$. In the SQUID setup (section \ref{sec:SQUID}) the compensation resistors were $R_\mathrm{C} = R = 500\,\Omega$ and were thermally well coupled to the mixing chamber. Typical off-balance signals of the Wheatstone bridges were on the order of $0.5\%$ (see also zero field values of Fig. \ref{fig:SignalComparison}a. As an additional effect of the cold compensation, the Johnson noise originating from the balancing resistors is greatly reduced especially at mixing chamber temperatures.\\

\subsection{Low Temperature Transformer Amplification}
\label{sec:LTT}

\begin{figure}[tb]
	\centering
		\includegraphics[width=0.95 \columnwidth]{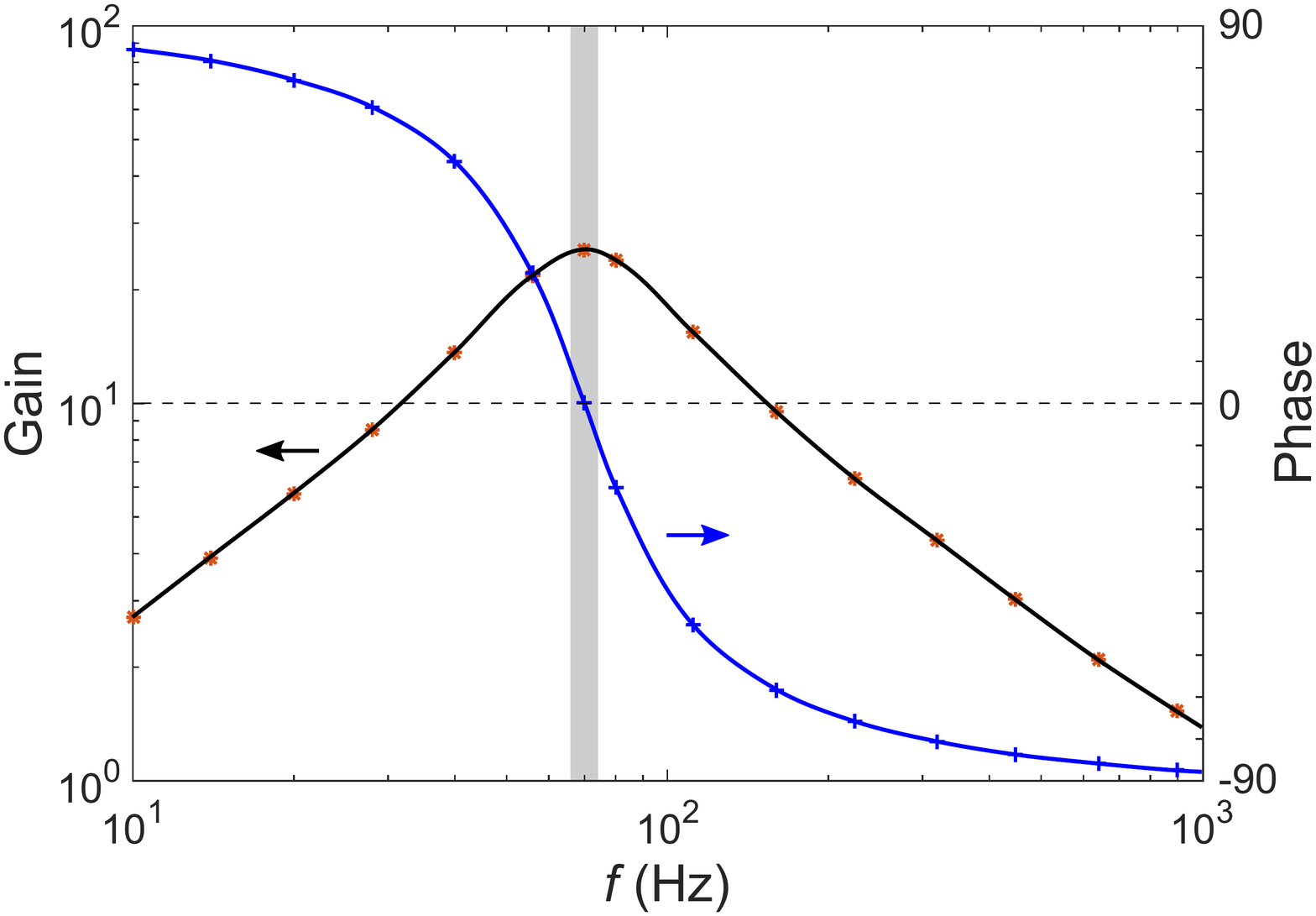}
	\caption{Frequency dependence of the low temperature transformer gain and phase for a LTT-m transformer with turns ratio $1:1000$ \cite{CMR} for a source impedance of $750\,\Omega$. The shaded region between $66$ and $74\,\mathrm{Hz}$ represents the useful bandwidth for which the gain is maximal and phase is minimal.}
	\label{fig:LTTGain}
\end{figure}

\noindent Due partly to the difficulties of using SQUID amplification in high magnetic fields, low temperature transformers were developed as an alternative low temperature amplification stage \cite{CMR}. In the most favorable circumstances of extremely low input impedance, they can give noise levels of below $1\,\mathrm{pV}$ \cite{Julian94}, but they are not well matched to the high source impedance of a piezo-lever. Figure \ref{fig:LTTGain} shows the associated gain and bandwidth issue. As can be seen the gain is strongly suppressed compared to its open circuit value of 1000. The sharp gain peak at 66 to $74\,\mathrm{Hz}$ limits the bandwidth of the LTT system. Thus cross talk between multiple simultaneous experiments amplified by LTTs can become an issue. Alternative room temperature tests with lower transformer ratios resulted in proportionally lower gains and marginally higher bandwidths, due to the tremendous impedance mismatch.\\
Nevertheless, for our single channel experiment on PdRhO$_2$, we attempted using lead-shielded LTTs \cite{CMR} to amplify the balanced voltage-signal. The amplified signal was directly fed into the PXI oscilloscope. By doing so a background noise level of $250\,\mathrm{nV/}\sqrt\mathrm{Hz}$ near the measurement frequency was achieved (see markers in Fig. \ref{fig:SpectralNoiseDensity}a). This is a factor of twelve above the bare noise of the unamplified measurement. At the same time the signal was amplified by a factor of 25 boosting the unity bandwidth SNR by a factor two compared to unamplified case (see also Fig. \ref{fig:SpectralNoiseDensity}).

\subsection{SQUID Amplification}
\label{sec:SQUID}

\begin{figure*}[tb]
	\centering
		\includegraphics[width=1.0\textwidth]{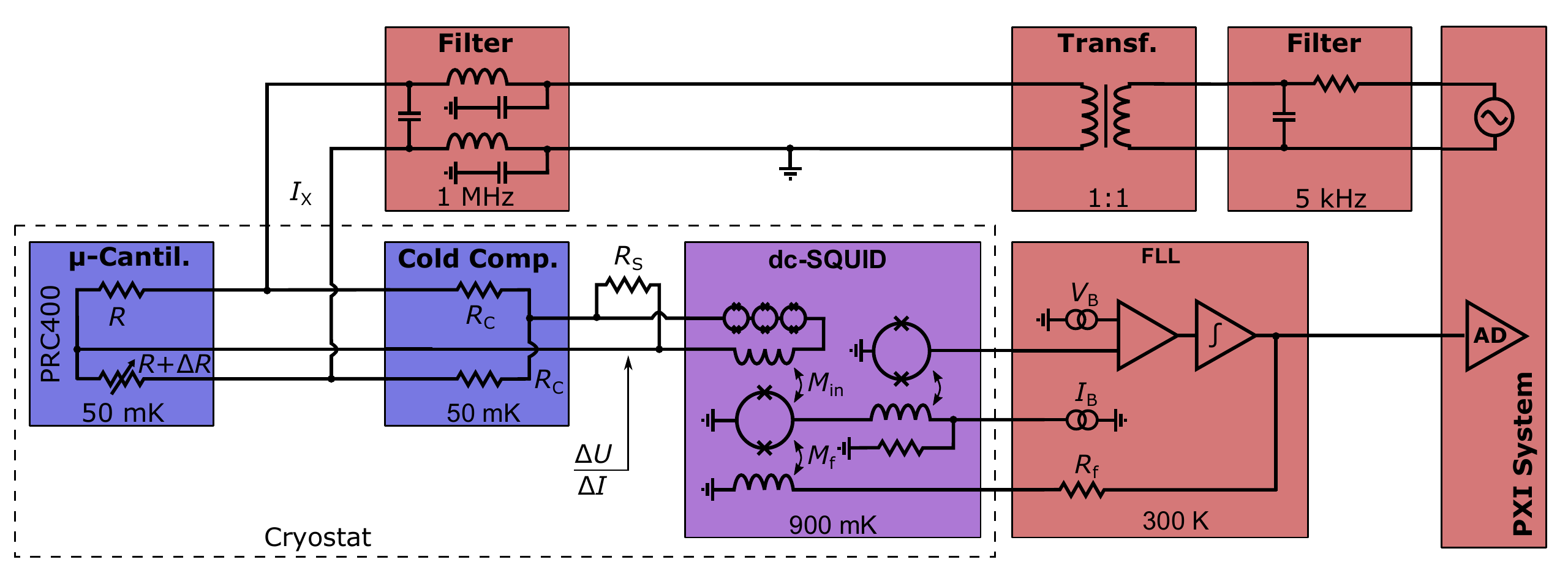}
	\caption{The schematic shows a simplified circuit diagram of the SQUID amplified magnetic torque measurement. The excitation current generated by the PXIe generator is filtered by a audio-transformer and low-pass filter. The off-balance of the Wheatstone bridge, consisting of the piezo-electric torque cantilevers ($R=500\,\Omega$) and balancing resistors ($R_\mathrm{C} = 500\,\Omega \hspace{1em} \mathrm{or} \hspace{1em} 1\,\mathrm{k}\Omega$), is amplified by the two-stage SQUID and digitized by a PXIe scope. The SQUID is controlled by the flux-locked-loop (FLL). The $R_\mathrm{S}=50\,\Omega$ shunt resistor forms a $4.4\,\mathrm{MHz}$ low pass filter with the SQUID input inductance.}
	\label{fig:CircuitDiagram}
\end{figure*}

\noindent In order to increase the SNR even further and to avoid the bandwidth limitation of LTTs, we replaced the transformers by a two-stage C6XXL116T SQUID manufactured by the Physikalisch Technische Bundesanstalt \cite{Drung07,PTB}. SQUIDs are highly sensitive, high gain current-to-voltage converters with an ultra low amplifier noise. The SQUID in our setup is enclosed in a niobium shield and mounted to the still, which is located in a field compensated region of our dilution refrigerator (see Fig. \ref{fig:Cryostat}). Biasing of the SQUID as well as the feed back loop are provided by a room temperature XXF Magnicon FLL \cite{Magnicon}. The final overall circuit diagram is shown in Fig. \ref{fig:CircuitDiagram}.\\
In going from LTTs to SQUID amplifiers, we noticed a general sensitivity of the SQUID to high frequency noise as is emitted by digital electronics, radio transmitters and switch-mode power supplies. Without proper grounding and filtering of cable shields, instruments and signal lines, we were not able to observe a $V\Phi$-characteristic of the input stage of our two-stage SQUID. In a first attempt, a 1:1 audio-transformer (bandwidth $20\,\mathrm{Hz}$ to $20\,\mathrm{kHz}$) and $5\,\mathrm{kHz}$ first order low-pass filter were installed to decouple the drive from the PXI ground and to suppress high frequency noise. An additional $1\,\mathrm{MHz}$ second order in-line low-pass filter was installed between the electronics rack and the cryostat to prevent differential and common mode high frequency noise from entering the cryostat \cite{Pobell}. A reference potential and high frequency drain were provided by an additional grounding point outside the cryostat. With these measures a stable operation of the SQUID was achieved.\\
Further improvements have been made by (a) installing a $R_\mathrm{S}=50\,\Omega$ shunt resistor across the SQUID input terminals, (b) moving the balancing circuit to the mixing chamber, and (c) changing the balancing resistors to $R_\mathrm{C}=500\,\Omega$. The shunt resistor and input inductance of the SQUID ($L_\mathrm{i} = 1.8\,\mu\mathrm{H}$) form a first-order low-pass filter with a cut-off frequency of $4.4\,\mathrm{MHz}$, prohibiting very and ultra-high frequency noise from entering the SQUID.\\
Contrary to the LTT, the excitation frequency of the SQUID amplified Wheatstone bridge can be chosen freely. Thus we minimize the noise level by selecting $571\,\mathrm{Hz}$, which is sufficiently far away from low frequency $1/f$-noise and harmonics of $50\,\mathrm{Hz}$.

\begin{figure}[tb!]
	\centering
	\includegraphics[width=0.9\columnwidth]{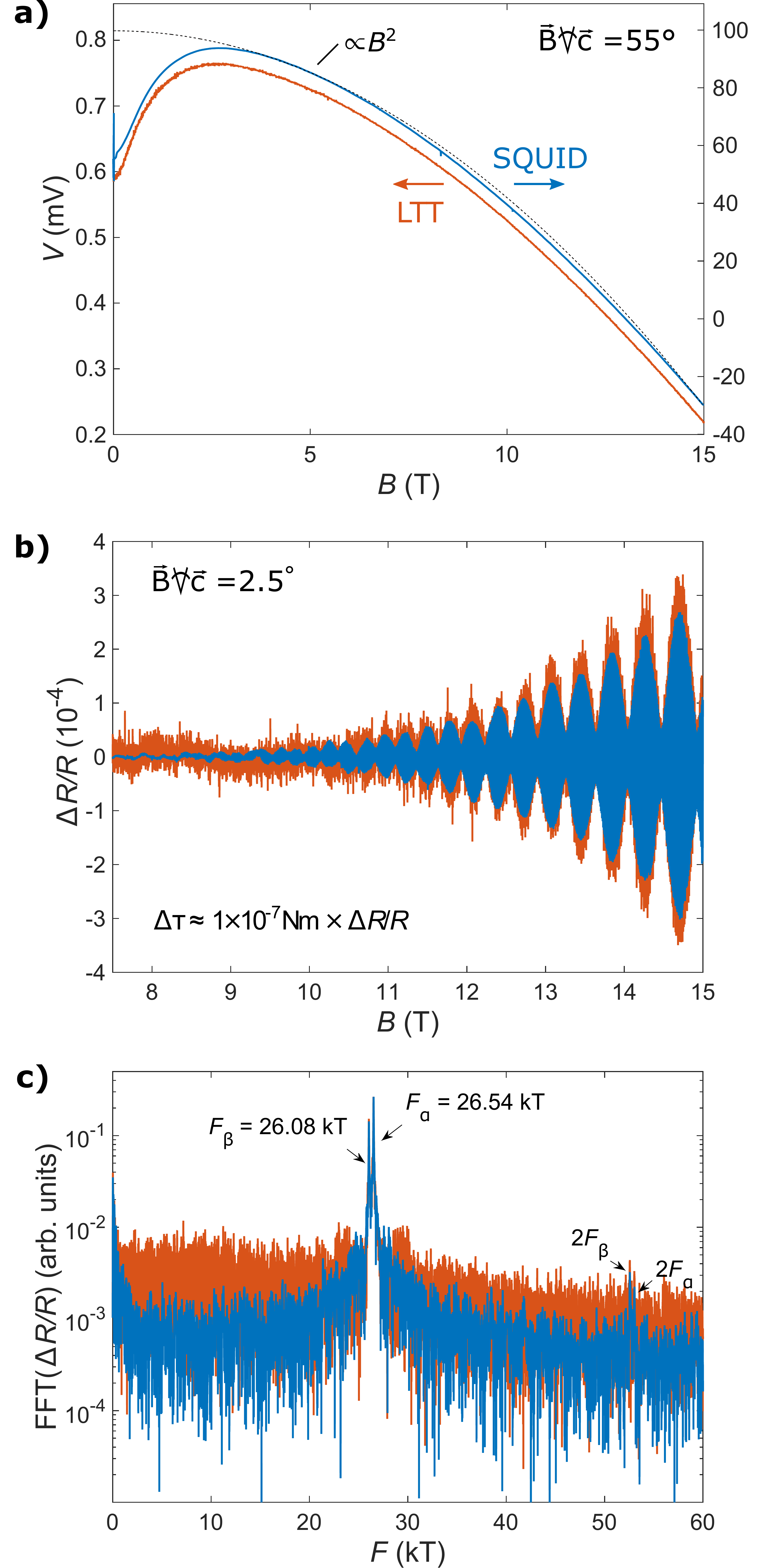}
	\caption{The graphs show \textbf{(a)} raw and \textbf{(b)} background subtracted magnetic torque data of a single PdRhO$_2$ crystal as well as the according de Haas-van Alphen spectrum \textbf{(c)} at an excitation current of $8\,\mu\mathrm{A}$ and a temperature of $100\,\mathrm{mK}$. LTT and SQUID amplified measurements were taken at a magnetic field angle of $55^\circ$ \textbf{(a)} and $2.5^\circ$ \textbf{(b,c)} within the (100)-plane with respect to the crystallographic c-axis. The dashed line in \textbf{(a)} shows a $B^2$ fit to the high field data. The relative resistance changes i.e. magnetic torque signals of \textbf{(b)} were calculated using Eqn. \ref{eqn:LTT} and \ref{eqn:SQUID}. Note for the SQUID amplified case, that the signal width below $10\,\mathrm{T}$ is determined by quantum oscillations and not by noise. In \textbf{c)}, both curves show the same amplitudes of the first and second order harmonics of the quantum oscillation frequencies.}
	\label{fig:SignalComparison}
\end{figure}

\section{Performance}
\label{sec:Performance}

\noindent We now evaluate the performance of our SQUID-amplified torque magnetometer based on a de Haas-van Alphen measurement on the delafossite PdRhO$_2$. For this a $300\times200\times50\mu\mathrm{m}$ sized PdRhO$_2$ single crystal was fixed on a PRC400 micro-cantilever and installed on our cryostat (see inset of Fig. \ref{fig:Cryostat}). Its magnetic torque was measured at a temperature of $(700\pm20)\,\mathrm{mK}$ in magnetic fields up to $15\,\mathrm{T}$. Data were taken during magnetic field down sweeps at a constant rate of $30\,\mathrm{mT/min}$. During a magnetic field sweep from 15 to $7.5\mathrm{T}$, i.e. in approximately four hours, we detect on the order of hundred flux jumps and on the order of ten integrator resets of our SQUID, showing the long term stability of the FLL. Whilst flux-jumps equilibrate on a time scale much shorter than our excitation frequencies and do not cause signal disturbances, integrator resets usually cause spikes in our lock-in signal. Most of these events can be traced back to broadband pulses in the main electrical power grid. Thus a post processing routine was applied to the data to remove these spikes.\\
As can be seen in Fig. \ref{fig:SignalComparison}a, at high fields the paramagnetic magnetization of PdRhO$_2$ induces a magnetic torque proportional to $B^2$, which is shown as the raw output voltage of the LTT and SQUID setup. Deviations from this behavior are likely due to saturating paramagnetic impurities at low fields. Both of the presented torque data were taken with the same sample and micro-cantilever.

Using the excitation current and gain of each setup, the relative resistance change $\Delta R/R$ of both Wheatstone bridges can be calculated. The measured gain of the LTT is $G_\mathrm{LTT}(66\,\mathrm{Hz})=25$ (see Fig. \ref{fig:LTTGain}), whilst the SQUID gain is determined by $G_\mathrm{S} =R_\mathrm{f}\times M_\mathrm{in}/M_\mathrm{f}$. For a feed-back resistor of $R_\mathrm{f}=10\,\mathrm{k}\Omega$ and mutual input $1/M_\mathrm{in}= 0.307\,\mu\mathrm{A}/\Phi_0$ and feed-back inductance $1/M_\mathrm{f}= 42.1\,\mu\mathrm{A}/\Phi_0$ the SQUID gain is $G_\mathrm{S}=1.37\,\mathrm{V/}\mu\mathrm{A}$.\\ The relative resistance change for a voltage read out Wheastone bridge (unamplified and LTT amplified) is:
\begin{eqnarray}
 \Delta R/R = 2\left(1/R+1/R_\mathrm{C}\right)\times\Delta V/I_X, 
\label{eqn:LTT}
\end{eqnarray}
which can be derived from a voltage divider on the active side under the assumption of $\Delta R \ll R$ (see Appendix A). For the SQUID amplified case, where the off balance current is measured, the formula is $\Delta I/I_X = \Delta R/(R+R_\mathrm{C})$ (see Appendix A). Taking into account that $R_\mathrm{C} \approx R$ in our SQUID setup, we obtain:
\begin{eqnarray}
 \Delta R/R = 2 \Delta I /I_X .
\label{eqn:SQUID} 
\end{eqnarray}

Figure \ref{fig:SignalComparison}b shows the $B^2$-background subtracted magnetic torque signals of the PdRhO$_2$ crystal given as $\Delta R/R$ for a magnetic field angle of $2.5^\circ$ with respect to the crystallographic c-axis within the (100)-plane. Dominant quantum oscillations and beating of the envelope function are visible due to the presence of two adjacent frequencies. A Fourier transform of the de Haas-van Alphen oscillations in $1/B$ can be found in Fig. \ref{fig:SignalComparison}c. Here two adjacent frequencies arise due to the warping of the quasi-two dimensional cylindrical Fermi surface along the $c$-direction. A full angular dependence and Fermi surface topography of PdRhO$_2$ can be found in \cite{Arnold17PRB}. As can be seen in Fig. \ref{fig:SignalComparison}b and c the SQUID amplified signal shows a clearly suppressed noise level compared to the LTT amplified case.\\
\begin{figure}[tb]
	\centering
		\includegraphics[width=1\columnwidth]{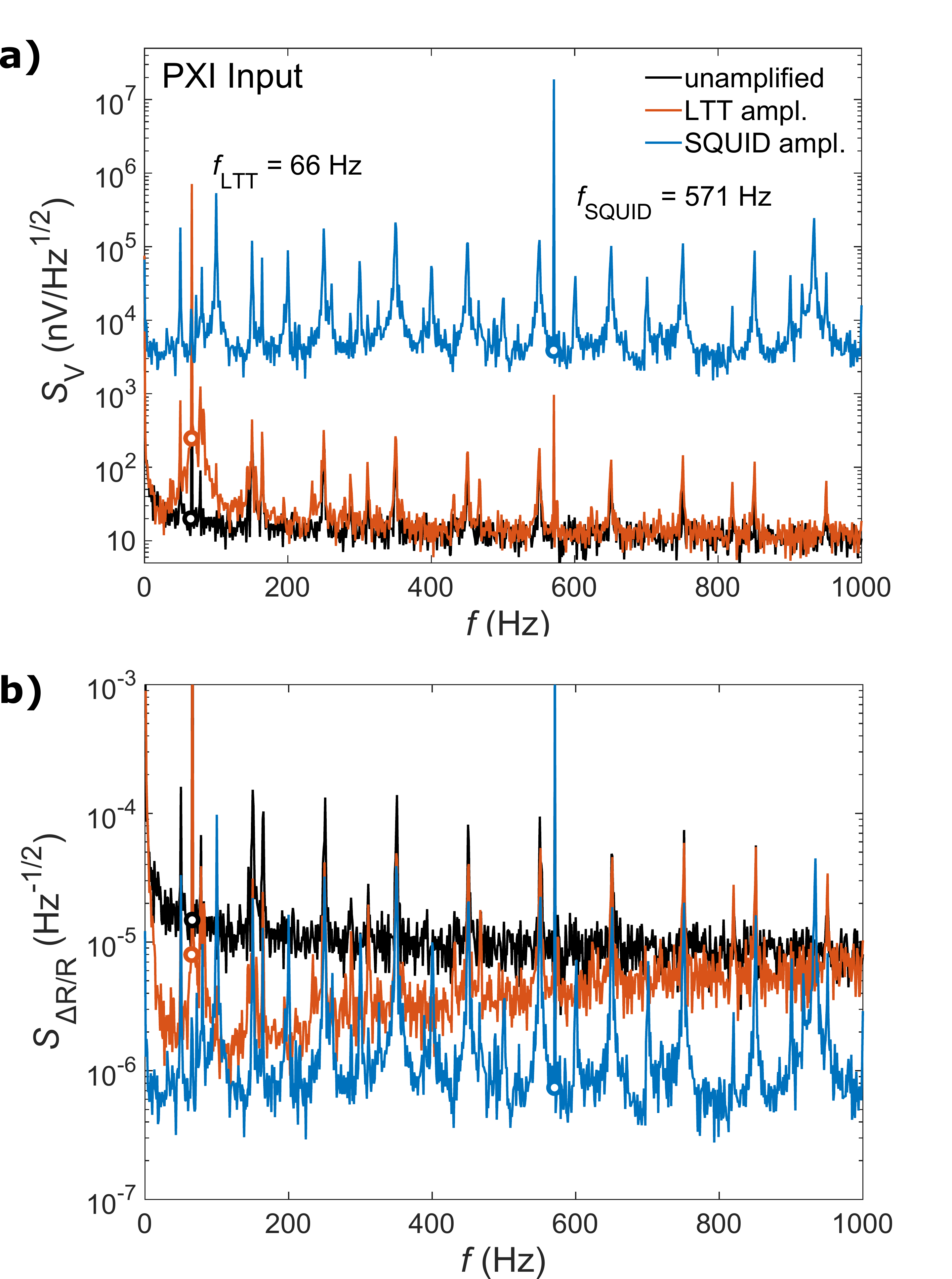}
	\caption{The graphs show the scope input spectral noise densities of an unamplified (black), LTT amplified (red) and SQUID amplified (blue) torque magnetometer at $15\,\mathrm{T}$.  \textbf{a)} shows the bare spectral noise density as recorded at the input of the PXI scope. The spectra were calculated from four subsequent voltage traces with a capturing time of $1\,\mathrm{s}$. Measurements were taken at an excitation current of $8\,\mu\mathrm{A}$ and frequencies of $66\,\mathrm{Hz}$ (unampl. and LTT ampl.) and $571\,\mathrm{Hz}$ (SQUID ampl.). Markers indicate the noise level at the respective excitation frequencies. Besides the signal peaks at the excitation frequencies, strong noise peaks at odd multiples of $50\,\mathrm{Hz}$ appear in the spectra, due to mechanical vibrations originating from vacuum pumps and electrical cross-talk. \textbf{b)} shows the spectral noise densities in units of the relative resistance changes $\Delta R / R$.}
	\label{fig:SpectralNoiseDensity}
\end{figure}

Figure \ref{fig:SpectralNoiseDensity}a shows the linear spectral noise densities $S_\mathrm{V}$ for all three methods. Here $S_\mathrm{V}(f)=\frac{1}{\sqrt{\theta}}\int_0^\theta V(t)e^{-2\pi i f t}dt$ where $\theta=1\,\mathrm{s}$ is the captured time of the truncated Fourier transform. Whilst the unamplified and SQUID amplified technique are broadband methods, the frequency using LTTs is limited to $(70\pm4)\,\mathrm{Hz}$. It should be noted that the signal peak heights in Fig.~\ref{fig:SpectralNoiseDensity} are generally not comparable between different experiments, as the zero torque off-balance signal is not zeroed in our experiment. The linear spectral noise densities near the given excitation frequencies are approximately $20\,\mathrm{nV/}\sqrt\mathrm{Hz}$ (unampl.), $250\,\mathrm{nV/}\sqrt\mathrm{Hz}$ (LTT ampl.) and $4\,\mu\mathrm{V/}\sqrt\mathrm{Hz}$ (SQUID ampl.) measured at the PXI scope and indicated as markers in Fig.~\ref{fig:SpectralNoiseDensity}a. In general, the spectral noise densities were found to be independent of excitation current, magnetic field and state of the magnet (persistent or driven mode). However, the noise level increased to $8\,\mu\mathrm{V/}\sqrt\mathrm{Hz}$ in the SQUID setup during fast field sweeps of 0.3~T/min.\\
In order to compare the resolving powers of the unamplified, LTT and SQUID amplified circuit, the resistance noise spectra are calculated by using the gains and applying Eqn. \ref{eqn:LTT} and \ref{eqn:SQUID} respectively, as before. The resulting noise spectra are shown in Fig.~\ref{fig:SpectralNoiseDensity}b. Since our post-aquisition digital lock-in uses a time constant of $\tau=1\,\mathrm{s}$ the linear noise densities are integrated over a bandwidth of $1\,\mathrm{Hz}$ around the measurement frequency. Close to the respective measurement frequencies, we obtain the root-mean-square resistance resolutions of $\Delta R/R=1.5\times10^{-5}$, $8\times10^{-6}$ and $7\times10^{-7}$ for the unamplified, LTT and SQUID amplified case (see markers in Fig.~\ref{fig:SpectralNoiseDensity}b). The SQUID amplified read-out achieves a ten to twenty times better resistance and magnetic torque resolution than conventional methods. This improved resolution can also be seen in the lower noise level (Fig. \ref{fig:SignalComparison}b and c) compared to the LTT amplified measurement. Note that the quoted resolutions are root-mean squares whereas Fig. \ref{fig:SignalComparison}b shows the absolute noise.\\
As can be seen in Eqn. \ref{eqn:LTT} and \ref{eqn:SQUID}, both resistance sensitivities scale with the excitation current and can therefore be made arbitrarily small by increasing the excitation current. However, power dissipation is a major issue for low and ultra-low temperature measurements. This is particularly true for the silicon based micro-cantilevers, with a low thermal conductance at low temperatures, when mounted in vacuum. The piezoelectric track of these cantilevers generates heat close to the sample, which is poorly thermalized to the platform. In Appendix B, we estimate the sample temperature based on the cantilever geometry and depending on the excitation current and platform temperature. As can be seen for an excitation current of $I_\mathrm{X}=8\mu\mathrm{A}$ ($I_\mathrm{R}=4\,\mu\mathrm{A}$ through the micro-cantilever), we reach a sample temperature of approximately $700\,\mathrm{mK}$.\\
Measurements to about $200\,\mathrm{mK}$ are possible by reducing the excitation current to approximately $100\,\mathrm{nA}$ (see Fig. \ref{fig:SampleTemperature}). Even lower temperatures at higher excitation currents might be achieved by mounting the micro-cantilever directly inside the mixing chamber or in a $^3$He submersion cell. In this case, the excitation current is only limited by the Kapitza resistance between the micro-cantilever and $^3$He liquid. However, much more sophisticated rotator mechanisms, such as piezo-electric rotators, would be required in order to study the angular dependence of the dHvA effect in these cells. 
\\
Following the theoretical torque calibration constant of $\tau = 1\times10^{-7}\,\mathrm{Nm}\times\Delta R/R$ \cite{Rossel96,McCollam11}, we obtain a torque resolution of $\Delta\tau = 7\times10^{-14}\,\mathrm{Nm}$ or equivalently magnetic moment resolution of $\Delta m=5\times10^{-15}\,\mathrm{J/T}$ at $B=15\,\mathrm{T}$ and $T=700\,\mathrm{mK}$. The latter is four orders of magnitude better than commercially available SQUID VSMs \cite{QD} at significantly lower base temperatures. Note that the resolution is inversely proportional to the excitation current, which itself is limited by the thermalization of the micro-cantilever and sample. Therefore, the resolution effectively decreases when lowering the sample temperature in our setup.\\
A disadvantage of the new level of precision, granted by the SQUID, is the general sensitivity to environmental fluctuations. Although the balancing resistors of the cold compensation have a temperature stability of better than $10^{-4}\,\mathrm{K}^{-1}$, minute changes of the $1\,\mathrm{K}$-pot or mixing chamber temperature are sufficient to induce slowly varying backgrounds in our measurements. Thus special care had to be taken to thermally decouple the balancing circuits from the $1\,\mathrm{K}$-pot and mixing chamber, whilst keeping them at a constantly low temperature.\\
At present the resolution of our setup is mainly limited by the Johnson noise of the $50\,\Omega$ shunt resistor across the SQUID terminals $(1.055\,\mathrm{pA/}\sqrt{\mathrm{Hz}})$, which is the dominant source of thermal noise and accounts for $30\%$ of the overall noise. Additionally, the output noise of the function generator $(50\,\mathrm{fA/}\sqrt{\mathrm{Hz}})$ and intrinsic SQUID noise $(200\,\mathrm{fA/}\sqrt{\mathrm{Hz}})$, account for another $10\,\%$ of the noise level of $4\,\mu\mathrm{V/}\sqrt{\mathrm{Hz}} / (1.37\,\mathrm{V}/\mathrm{\mu A}) = 2.9\,\mathrm{pA}/\sqrt{\mathrm{Hz}}$ (for further details see Appendix C). We hypothesize, that the remaining $60\,\%$ of the observed noise originate from mechanical pick-up and random-telegraph-noise within the micro-cantilevers \cite{Rossel96} as well as noise entering the FLL wiring. 

\section{Conclusion}

\noindent In summary, we have successfully developed and built a new highly sensitive, ultra-low noise torque magnetometer for sub-millimeter sized samples suitable for high magnetic fields and low temperatures. The magnetometer is based on a standard piezo-electric micro-cantilever and utilizes a two-stage SQUID as the null-detector of a cold Wheatstone bridge. We were able to demonstrate its performance in a de Haas-van Alphen experiment of the metallic delafossite PdRhO$_2$ down to $700\,\mathrm{mK}$ and up to $15\,\mathrm{T}$ and achieved a torque resolution $\Delta \tau=7\times10^{-14}\,\mathrm{Nm}$ at an excitation current of $8\,\mu\mathrm{A}$.\\
This is the first successful use of a SQUID in a resistance measurement of such high resolution up to $15\,\mathrm{T}$. Comparing our setup to conventional low temperature techniques, we were able to show that SQUID amplification offers up to one order of magnitude higher resolution than unamplified and low temperature transformer amplified Wheatstone bridges.\\
Due to the general applicability of balanced bridge circuits to highly sensitive resistance, inductance and capacitance measurements, we would like to point out the possibility of applying SQUID amplified read-outs to many electrical and thermal transport, ac-susceptibility, heat-capacity as well as thermal expansion and magnetostriction experiments, even in magnetic fields up to $15\,\mathrm{T}$. 

\section{Acknowledgments}

The authors would like to acknowledge J. Saunders and A. Casey \textit{et al.} at the Royal Holloway University of London for preceding fruitful discussions and ideas as well as P.-J. Zermatten for technical advice and discussion of the manuscript. Furthermore we would like to thank the Max-Planck Society and Deutsche Forschungsgemeinschaft, project "Fermi-surface topology and emergence of novel electronic states in strongly correlated electron systems", for their financial support.

\appendix

\section{Resistance resolution of voltage and current read-out Wheatstone bridges}

\begin{figure}[tb]
	\centering
		\includegraphics[width=0.95\columnwidth]{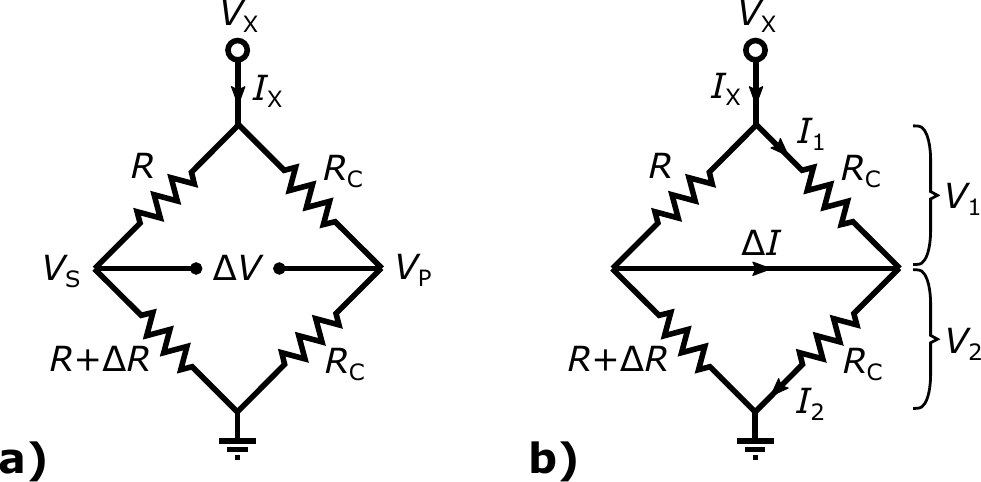}
	\caption{\textbf{a)} Schematic circuit diagram of voltage read-out and \textbf{b)} of a current read-out Wheatstone bridge.}
	\label{fig:WheatstoneBridge}
\end{figure}

\noindent In this appendix, we derive the resistance resolution for voltage and current read out Wheatstone bridges. For simplicity we assume the circuit diagrams of Fig. \ref{fig:WheatstoneBridge}.\\
The voltage off-balance in a Wheatstone bridge $\Delta V = V_\mathrm{S} - V_\mathrm{P}$ is the difference between the sensing and passive side of the Wheatstone bridge. For large voltmeter input impedances and equal compensation resistors $R_\mathrm{C}$ these are given by:
\begin{eqnarray}
V_\mathrm{S} &=& \frac{R+\Delta R}{2R+\Delta R} V_\mathrm{X} \hspace{1em}\mathrm{and}\hspace{1em} V_\mathrm{P} = \frac{1}{2}V_\mathrm{X}.
\end{eqnarray}
Combining them, one obtains:
\begin{eqnarray}
\Delta V &=& \frac{\Delta R}{4R+2\Delta R}V_\mathrm{X}.
\end{eqnarray}
Under the assumption of small resistance changes, i.e. $\Delta R \ll R$, the equation simplifies to:
\begin{eqnarray}
\frac{\Delta R}{R} &\approx& 4\frac{\Delta V}{V_X} = 2\frac{\Delta V}{I_X}\left(\frac{1}{R}+\frac{1}{R_C}\right)
\end{eqnarray}
In a current read out bridge, the off-balance current $\Delta I = I_{2}-I_{1}$. These can be calculated from the voltage drop across the balancing resistors:
\begin{eqnarray}
I_{1} = \frac{V_1}{R_C}\hspace{1em},\hspace{1em}I_{2} = \frac{V_2}{R_C}.
\end{eqnarray}
As the superconducting SQUID input coil presents a negligible impedance at low frequency, an ideal short between the sensing and passive side can be assumed. Thus $R$ and $R_C$ form parallel networks and one obtains:
\begin{eqnarray}
V_1 &=& I_X\left(\frac{1}{R}+\frac{1}{R_C}\right)^{-1},\\
V_2 &=& I_X\left(\frac{1}{R+\Delta R}+\frac{1}{R_C}\right)^{-1}.
\end{eqnarray}
Combining these with the above leads to:
\begin{eqnarray}
\Delta I &=& \frac{I_X}{R_C}\left[\frac{RR_C+\Delta RR_C}{R+\Delta R+R_C}-\frac{RR_C}{R+R_C}\right].
\end{eqnarray}
And by assuming $\Delta R \ll R$ this simplifies to:
\begin{eqnarray}
\frac{\Delta I}{I_X}&=&\frac{\Delta R}{R+R_C}.
\end{eqnarray}

\section{Sample Temperature}

\begin{figure}[tb]
	\centering
		\includegraphics[width=1\columnwidth]{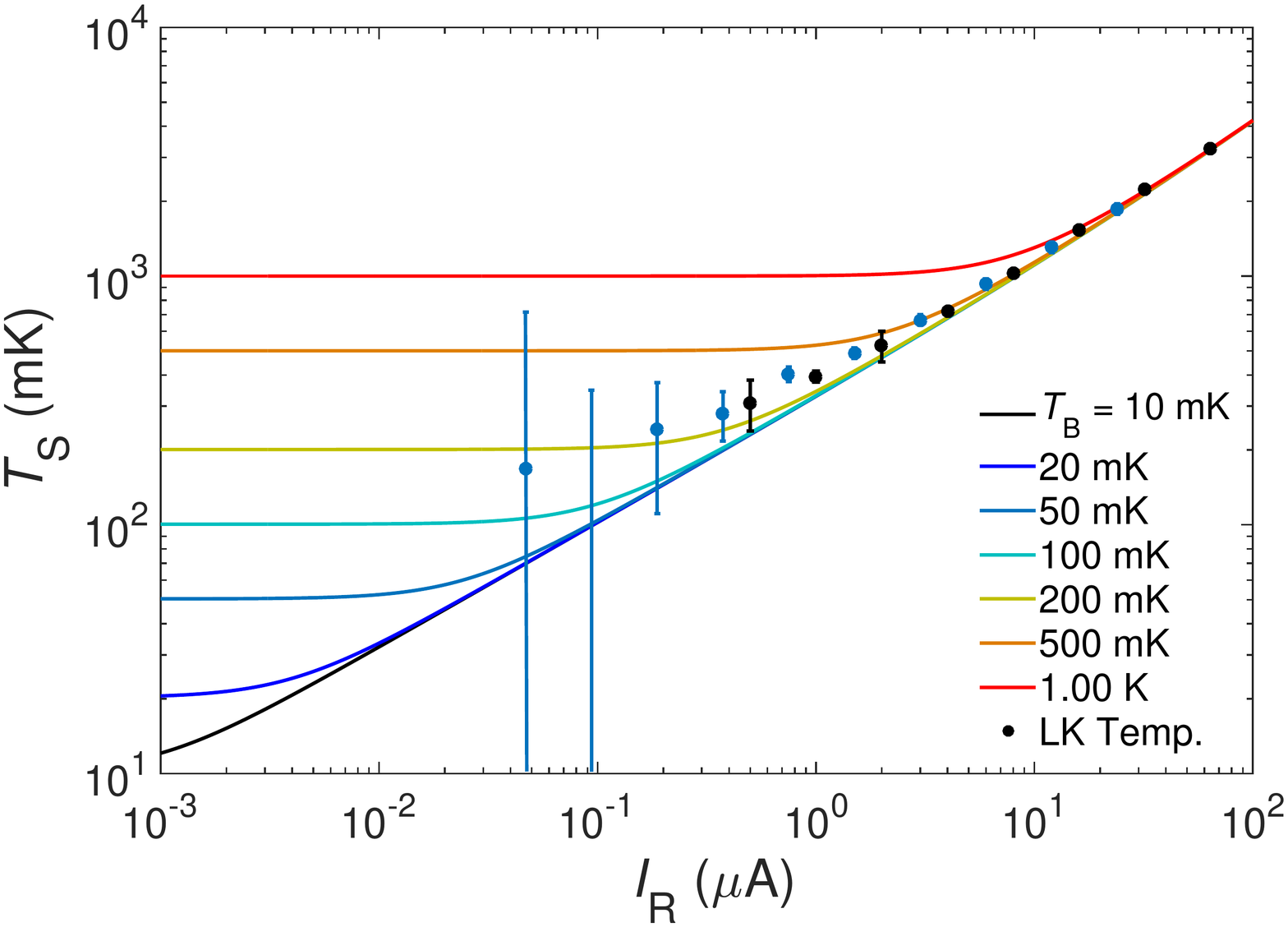}
	\caption{Theoretical and experimental sample temperatures as a function of excitation current and base temperatures for a PRC400 micro-cantilever \cite{Hitachi}. Dots mark experimental temperatures determined from fits of the Lifshitz-Kosevich temperature reduction term to the quantum oscillation amplitude of Sr$_2$RuO$_4$ (blue) and PdRhO$_2$ (black). Lines are calculated temperatures as described in the text.}
	\label{fig:SampleTemperature}
\end{figure}

\begin{figure}[tb]
	\centering
		\includegraphics[width=0.7\columnwidth]{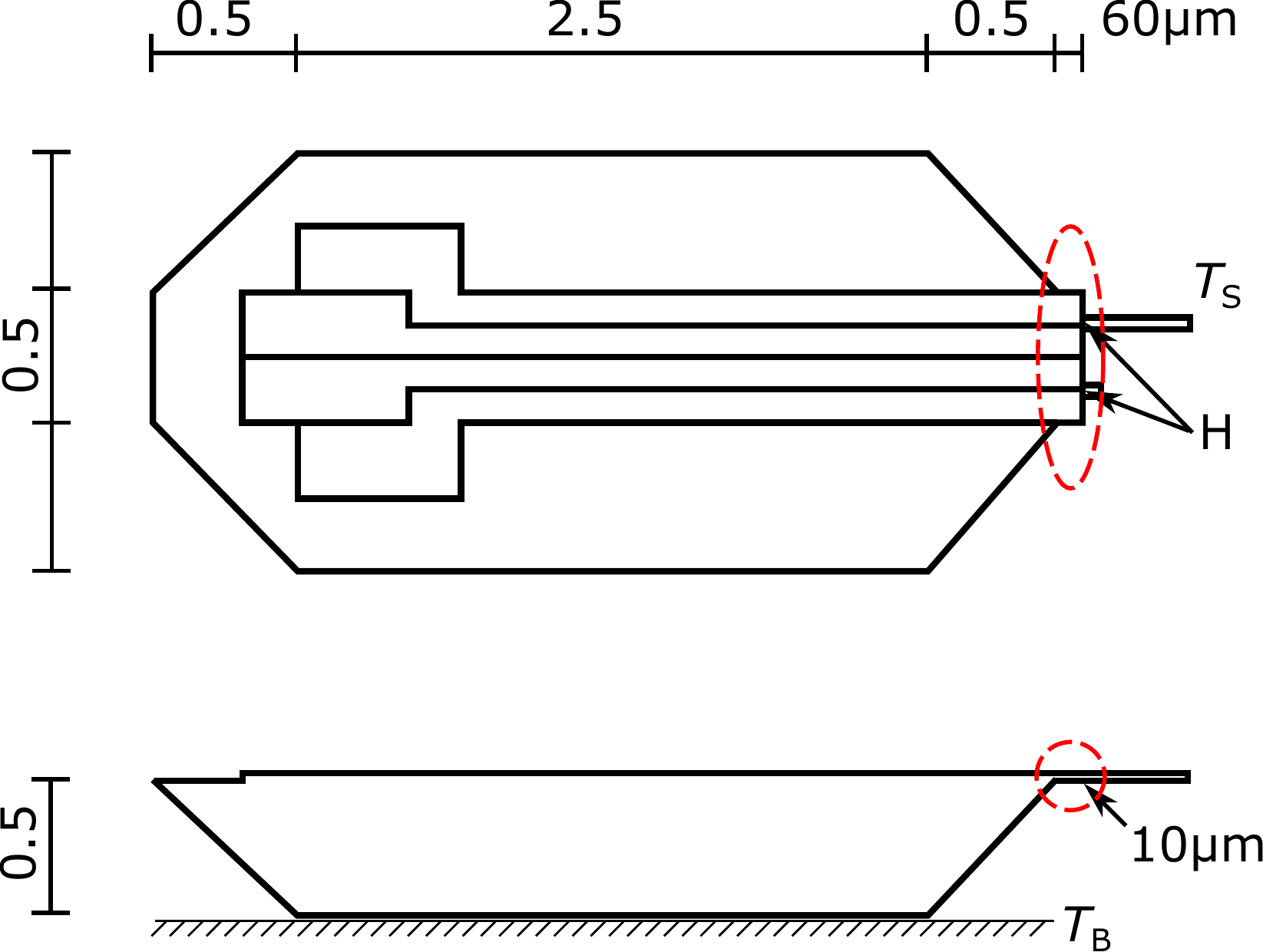}
	\caption{Microcantilever geometry used for thermalization calculations. H marks the position of the heating piezo elements, $T_\mathrm{B}$ and $T_\mathrm{S}$ are the base and sample temperature respectively. The part encircled in red is the thermal bottle neck that accounts for the majority of the thermal gradient.}
	\label{fig:CantileverSchematic}
\end{figure}

\noindent The sample temperature $T_\mathrm{S}$ was studied as a function of the excitation current $I_\mathrm{R}$ for our PRC400 piezo-electric micro-cantilever \cite{Hitachi}, which is mounted in vacuum and heat sunk through a thin layer of Apiezon N grease to a thermal bath of temperature $T_B$. $T_\mathrm{S}$ was determined by measuring the excitation current dependence of the quantum oscillation amplitude of a Sr$_2$RuO$_4$ and PdRhO$_2$ crystal (see solid points in Fig. \ref{fig:SampleTemperature}) for $T_{\mathrm{B}} = 50\,\mathrm{mK}$. Note that $T_{\mathrm{B}}$ increased when the current was higher than $10\,\mu$A. The sample temperatures were calculated by applying the Lifshitz-Kosevich temperature reduction term to the observed amplitude suppression. The cyclotron masses of both materials were taken from \cite{Bergemann00,Arnold17PRB}. The error bars at low excitation currents are dominated by the resolution of the quantum oscillation amplitude and the flatness of the amplitude versus temperature curve. At high currents the error on the effective masses and the estimated zero-temperature quantum oscillation amplitude are predominant.\\
\noindent We compare the experimental result with a geometrical model where the heat is conducted through the silicon cantilever and a thin layer of Apiezon N grease ($\kappa_\mathrm{Ap}(T)=10^{-5}\,\mathrm{Wcm}^{-1}\mathrm{K}^{-3} \times T^2$)\cite{Pobell}. The effective geometry is shown in Fig. \ref{fig:CantileverSchematic}. Thermal boundary resistances were ignored. Due to their poor aspect ratio and correspondingly high thermal resistance, the gold leads connecting to the piezo-resistive tracks do not contribute to the thermalization of the cantilever and could also be ignored. The bottle-neck of the thermal path at low temperature is the thinned down end of the silicon chip which the micro-cantilevers are attached to. The result is shown as lines in Fig. \ref{fig:SampleTemperature}. The model fits nicely for excitation currents $I_{\mathrm{R}} > 3\,\mu \mathrm{A}$ when the thermal conductivity of the epitaxial silicon cantilevers is taken to be three orders of magnitude lower than the literature value\cite{Pobell} for bulk crystalline silicon of $\kappa = 2 \times 10^{-1} \,\mathrm{W\,cm}^{-1}\,\mathrm{K}^{-4}\times T^3$.
Based on the literature value and without a direct measurement using quantum oscillations of a sample with higher effective cyclotron mass, the sample temperature for $I_{\mathrm{R}} = 4\,\mu\mathrm{A}$ would be underestimated to be around $100\,\mathrm{mK}$. For lower excitation currents the sample temperature seems to be higher than in our model, levelling off at roughly $200\,\mathrm{mK}$. The difference might be ascribed to additional heating induced by rf noise not included in our model.

\section{Noise Sources}

\noindent The Johnson noise at the SQUID input arises mainly from resistors in the Wheatstone bridge. The influence of external room temperature resistors is lowered by the almost perfect balancing of the Wheatstone bridge. As can be seen in Fig. \ref{fig:WheatstoneBridge}b, the Johnson noise of the piezoelectric track and balancing resistor induce a current noise in the upper and lower branch of the Wheatstone bridge. The thermal noise power per unit bandwidth of each of these resistors is given by:
\begin{eqnarray}
 S^2_\mathrm{V}=4k_\mathrm{B}TR.
\end{eqnarray}
Note that in resistor networks the total noise is given by the sum of the mean-square noises due to the uncorrelated nature of the individual noise sources. Thus, in the low frequency limit ($f\ll f_{-3dB} \approx 70\,\mathrm{MHz}$), where the input impedance of the SQUID is ignored, the arising noise current from each branch through the SQUID is given by:
\begin{eqnarray}
S_\mathrm{I} = \frac{\sqrt{S^2_\mathrm{V}(R)+S^2_\mathrm{V}(R_\mathrm{C})}}{R+R_\mathrm{C}},
\end{eqnarray}
where the current noise $S_\mathrm{I}$ is limited by the total serial resistance $R+R_\mathrm{C}$ of each branch. Summing over both branches, leads to the total noise current at the SQUID input:
\begin{eqnarray}
S_\mathrm{I} = \frac{\sqrt{2\times(4k_\mathrm{B}TR+4k_\mathrm{B}T_\mathrm{C}R_\mathrm{C})}}{R+R_\mathrm{C}}.
\end{eqnarray}
In the present setup $R=R_\mathrm{C}=500\,\Omega$,  $T= 700\,\mathrm{mK}$, and the temperature of the compensation resistors $T_\mathrm{C}=100\,\mathrm{mK}$, leading to an over all Johnson-noise level of balancing resistors of $S_\mathrm{I}=200\,\mathrm{fA/}\sqrt{\mathrm{Hz}}$. The shunt resistor $R_\mathrm{S}=50\,\Omega$ at $T=1\,\mathrm{K}$ on the other hand gives rise to a Johnson noise of $1.05\,\mathrm{pA/}\sqrt{\mathrm{Hz}}$ and is clearly dominating the noise of the balancing resistors.\\
Additional noise arises from the output of the PXIe-4463 function generator. Following the manufacturers data sheet, the according output noise level is $60\,\mathrm{nV/}\sqrt{\mathrm{Hz}}$ at $100\,\mathrm{Hz}$. Taking into account the total impedance of the circuit and a Wheatstone bridge off-balance of $0.5\%$, this amounts to a noise current of $S_\mathrm{I}=50\,\mathrm{fA/}\sqrt{\mathrm{Hz}}$ at the SQUID input.

\end{document}